\documentclass[superscriptaddress,showpacs,amsmath,amssymb,aps,prd,twocolumn,floatfix,nofootinbib]{revtex4-1}

\usepackage{graphicx}
\graphicspath{{figures/}}

\usepackage{dcolumn}
\usepackage{bm}
\usepackage{color}
\usepackage{amsmath}
\usepackage{amssymb}
\usepackage{bbm}
\usepackage{subfigure}
\usepackage{hyperref}
\usepackage{hhline}
\usepackage[percent]{overpic}
\usepackage{layouts}
\usepackage{float}
\usepackage{comment}
\usepackage{graphicx}
\usepackage{adjustbox}
\usepackage{placeins}
\usepackage{ amssymb }
\usepackage{siunitx}
\usepackage{caption}

\newcommand{\sssnet}{\textsc{SparseSSNet}}
\usepackage{lineno}
\usepackage{multirow}
\usepackage{float}

\begin{document}

\title{Semantic Segmentation with a Sparse Convolutional Neural Network for Event Reconstruction in MicroBooNE}

\newcommand{\Bern}{Universit{\"a}t Bern, Bern CH-3012, Switzerland}
\newcommand{\BNL}{Brookhaven National Laboratory (BNL), Upton, NY, 11973, USA}
\newcommand{\UCSB}{University of California, Santa Barbara, CA, 93106, USA}
\newcommand{\Cambridge}{University of Cambridge, Cambridge CB3 0HE, United Kingdom}
\newcommand{\StKates}{St. Catherine University, Saint Paul, MN 55105, USA}
\newcommand{\CIEMAT}{Centro de Investigaciones Energ\'{e}ticas, Medioambientales y Tecnol\'{o}gicas (CIEMAT), Madrid E-28040, Spain}
\newcommand{\Chicago}{University of Chicago, Chicago, IL, 60637, USA}
\newcommand{\Cincinnati}{University of Cincinnati, Cincinnati, OH, 45221, USA}
\newcommand{\CSU}{Colorado State University, Fort Collins, CO, 80523, USA}
\newcommand{\Columbia}{Columbia University, New York, NY, 10027, USA}
\newcommand{\FNAL}{Fermi National Accelerator Laboratory (FNAL), Batavia, IL 60510, USA}
\newcommand{\Granada}{Universidad de Granada, Granada E-18071, Spain}
\newcommand{\Harvard}{Harvard University, Cambridge, MA 02138, USA}
\newcommand{\IIT}{Illinois Institute of Technology (IIT), Chicago, IL 60616, USA}
\newcommand{\KSU}{Kansas State University (KSU), Manhattan, KS, 66506, USA}
\newcommand{\Lancaster}{Lancaster University, Lancaster LA1 4YW, United Kingdom}
\newcommand{\LANL}{Los Alamos National Laboratory (LANL), Los Alamos, NM, 87545, USA}
\newcommand{\Manchester}{The University of Manchester, Manchester M13 9PL, United Kingdom}
\newcommand{\MIT}{Massachusetts Institute of Technology (MIT), Cambridge, MA, 02139, USA}
\newcommand{\Michigan}{University of Michigan, Ann Arbor, MI, 48109, USA}
\newcommand{\Minnesota}{University of Minnesota, Minneapolis, MN, 55455, USA}
\newcommand{\NMSU}{New Mexico State University (NMSU), Las Cruces, NM, 88003, USA}
\newcommand{\Otterbein}{Otterbein University, Westerville, OH, 43081, USA}
\newcommand{\Oxford}{University of Oxford, Oxford OX1 3RH, United Kingdom}
\newcommand{\PNNL}{Pacific Northwest National Laboratory (PNNL), Richland, WA, 99352, USA}
\newcommand{\Pitt}{University of Pittsburgh, Pittsburgh, PA, 15260, USA}
\newcommand{\Rutgers}{Rutgers University, Piscataway, NJ, 08854, USA}
\newcommand{\StMarys}{Saint Mary's University of Minnesota, Winona, MN, 55987, USA}
\newcommand{\SLAC}{SLAC National Accelerator Laboratory, Menlo Park, CA, 94025, USA}
\newcommand{\SDSMT}{South Dakota School of Mines and Technology (SDSMT), Rapid City, SD, 57701, USA}
\newcommand{\Maine}{University of Southern Maine, Portland, ME, 04104, USA}
\newcommand{\Syracuse}{Syracuse University, Syracuse, NY, 13244, USA}
\newcommand{\TelAviv}{Tel Aviv University, Tel Aviv, Israel, 69978}
\newcommand{\Tennessee}{University of Tennessee, Knoxville, TN, 37996, USA}
\newcommand{\UTA}{University of Texas, Arlington, TX, 76019, USA}
\newcommand{\Tufts}{Tufts University, Medford, MA, 02155, USA}
\newcommand{\VTech}{Center for Neutrino Physics, Virginia Tech, Blacksburg, VA, 24061, USA}
\newcommand{\Warwick}{University of Warwick, Coventry CV4 7AL, United Kingdom}
\newcommand{\Yale}{Wright Laboratory, Department of Physics, Yale University, New Haven, CT, 06520, USA}

\affiliation{\Bern}
\affiliation{\BNL}
\affiliation{\UCSB}
\affiliation{\Cambridge}
\affiliation{\StKates}
\affiliation{\CIEMAT}
\affiliation{\Chicago}
\affiliation{\Cincinnati}
\affiliation{\CSU}
\affiliation{\Columbia}
\affiliation{\FNAL}
\affiliation{\Granada}
\affiliation{\Harvard}
\affiliation{\IIT}
\affiliation{\KSU}
\affiliation{\Lancaster}
\affiliation{\LANL}
\affiliation{\Manchester}
\affiliation{\MIT}
\affiliation{\Michigan}
\affiliation{\Minnesota}
\affiliation{\NMSU}
\affiliation{\Otterbein}
\affiliation{\Oxford}
\affiliation{\PNNL}
\affiliation{\Pitt}
\affiliation{\Rutgers}
\affiliation{\StMarys}
\affiliation{\SLAC}
\affiliation{\SDSMT}
\affiliation{\Maine}
\affiliation{\Syracuse}
\affiliation{\TelAviv}
\affiliation{\Tennessee}
\affiliation{\UTA}
\affiliation{\Tufts}
\affiliation{\VTech}
\affiliation{\Warwick}
\affiliation{\Yale}

\author{P.~Abratenko} \affiliation{\Tufts} 
\author{M.~Alrashed} \affiliation{\KSU}
\author{R.~An} \affiliation{\IIT}
\author{J.~Anthony} \affiliation{\Cambridge}
\author{J.~Asaadi} \affiliation{\UTA}
\author{A.~Ashkenazi} \affiliation{\MIT}\affiliation{\TelAviv}
\author{S.~Balasubramanian} \affiliation{\Yale}
\author{B.~Baller} \affiliation{\FNAL}
\author{C.~Barnes} \affiliation{\Michigan}
\author{G.~Barr} \affiliation{\Oxford}
% ...

\collaboration{The MicroBooNE Collaboration}
\thanks{microboone\_info@fnal.gov}\noaffiliation


\begin{abstract}

We present the performance of a semantic segmentation network, \textsc{SparseSSNet}, that provides pixel-level classification of MicroBooNE data. The MicroBooNE experiment employs a liquid argon time projection chamber for the study of neutrino properties and interactions. \sssnet\ is a submanifold sparse convolutional neural network, which provides the initial machine learning based algorithm utilized in one of MicroBooNE’s $\nu_e$-appearance oscillation analyses. The network is trained to categorize pixels into five classes, which are re-classified into two classes more relevant to the current analysis. The output of \sssnet\ is a key input in further analysis steps. This technique, used for the first time in liquid argon time projection chambers data and is an improvement compared to a previously used convolutional neural network, both in accuracy and computing resource utilization. The accuracy achieved on the test sample is $\geq 99\%$. For full neutrino interaction simulations, the time for processing one image is $\approx$\,0.5\,sec, the memory usage is at 1 GB level, which allows utilization of most typical CPU worker machine. 
\end{abstract}

\maketitle

\section{Introduction}
\label{sec:intro}
The primary goal of the MicroBooNE experiment is to search for electron-like events, specifically in the kinematic region where an anomaly was reported by the MiniBooNE experiment~\cite{Aguilar-Arevalo:2018gpe}. The MiniBooNE experiment observed an excess with respect to their background predictions in electron neutrino events below 500\,MeV. This excess is often referred to as the MiniBooNE low energy excess (LEE). Four independent analyses from MicroBooNE are targeted at explaining this excess. The analyses differ in reconstruction techniques and in their approach of targeting different signal topologies. Specifically, the deep learning (DL) LEE analysis uses a combination of machine learning algorithms~\cite{Abratenko:2020pbp} and traditional tracking tools~\cite{Abratenko:2020wum}. As charged current quasi-elastic (CCQE) is the dominant cross section in the LEE energy range, the approach adopted by the DL-LEE analysis is to study high-purity data samples of CCQE $\nu_e$ and $\nu_\mu$ interactions. The topologies of these interactions are much simpler than other interaction types since they manifest in most cases as one lepton and one proton ($1l1p$), where the lepton is either an electron ($1e1p$) or a muon ($1\mu1p$) for $\nu_e$ or $\nu_\mu$ interactions, respectively. Focusing on these topologies allows an easier and more precise event selection than trying to select topologies with neutral particles in final states or hadronic interactions.

The MicroBooNE experiment has been collecting data since 2015 and is part of the Short-Baseline Neutrino (SBN) program~\cite{Antonello:2015lea} operating at Fermi National Accelerator Laboratory (FNAL) along the Booster Neutrino Beamline (BNB)~\cite{AguilarArevalo:2008yp}. The detector~\cite{Acciarri:2017sde} is a 10.4\,m long, 2.6\,m wide, and 2.3\,m high liquid argon time projection chamber (LArTPC), consisting of 170 tons (85 tons in the active volume). The readout time window is 4.8\,ms and is digitized into 9600 readout time ticks. Upon neutrino-argon interaction, various particles are produced depending on the interaction channel. The final state charged particles produce prompt scintillation light as well as ionization electrons along their paths in the liquid argon. The light is detected by an array of 32 Hamamatsu 5912-02MOD photomultiplier tubes (PMTs), while the ionization electrons drift in an electric field of 273\,V/cm towards the readout wire planes leading to a drift time of 2.3\,ms for the maximal distance. There are three wire planes: two induction planes (referred to as U and V) consisting of 2400 wires each and one collection plane (referred to as Y) consisting of 3456 wires. The induction plane wires are aligned at $\pm 60^\circ$ with respect to those of the collection plane, while the collection plane wires are vertical. The distance between two adjacent wires (on the same plane) is 3mm for all planes.

In LArTPCs the ionization pattern generated by a proton or a muon is a straight line referred to as a track~\cite{Abratenko:2019jqo}. It originates at the interaction point and extends to the point where the particle does one or several of the following: loses all its energy and comes to a stop; interacts again; decays; or exits the detector active volume. The most significant difference between a proton track and a muon track is in the energy deposition per unit length (dE/dx). The electromagnetic shower pattern generated by an electron (E$\geq39$\,MeV~\cite{Adams:2019law}) has a richer topology that is similar to a tree with many random branches. This is due to the electron losing energy to ionization as well as stochastically emitting photons. The emitted photons produce an electron-positron pair or Compton scatter to produce electrons. At this stage, the same processes occur again until all energy is deposited in the detector or the shower exceeds the detector boundaries. This cascade generates a pattern referred to as a shower~\cite{Adams:2019law}. The lower the electromagnetic shower energy is, the less branches are created and its topology becomes less distinct from a track. Correctly classifying the signature generated by charged particles in the detector as a shower or a track is a key ingredient of the DL-LEE analysis.

In the MicroBooNE detector, the data from the wires are retrieved as waveforms. As a first step, the waveforms are be subjected to signal processing that reconstructs the original ionization charge and zero-suppresses non-signal regions (see~\cite{Acciarri:2017sde}).

In the DL-LEE analysis, the data is represented as a set of three two-dimensional images (one for each wire plane), with wire number plotted along the x-axis and drift time plotted along the y-axis. The intensity of each ``pixel'', measured in pixel intensity units (PIUs), gives a measure of the number of ionization electrons arriving at the corresponding location and time. Along the time axis, the waveform is integrated over six TPC time-ticks (3$\mu$s). Thus the effective size of each pixel is 3.3\,mm along the y-axis and 3\,mm along the x-axis. After signal-processing, the resulting image will contain sparse regions of interest (ROIs) that are the charge-signals from the interaction. Pixels outside of these ROIs, are set to zero. To assure images from all wire planes are the same size, the images from the induction planes are padded with zero value pixels for wires 2401-3456 yielding a final images size of 3456~$\times$~1008.

Convolutional neural networks (CNNs) are the state of the art algorithms for solving many problems in image processing~\cite{DL}. In recent years machine learning techniques in general, and specifically CNNs, have seen many applications in physics~\cite{Carleo:2019ptp}. Particularly in the field of neutrino physics, many data analyses exploit the power of CNNs for various tasks such as event classification, background rejection, energy reconstruction, and more~\cite{Aurisano:2016jvx,Acciarri:2016ryt,Adams:2018bvi,Kekic:2020cne,Delaquis:2018zqi}. As the MicroBooNE data can be represented by sets of images, it is natural to exploit the excellent performances of CNNs. Recently, an implementation of CNNs oriented at sparse data sets, named Submanifold Sparse Convolutional Networks (SSCNs), was proposed by the Facebook AI team~\cite{DBLP:journals/corr/GrahamM17,Graham_2018_CVPR} and has drawn much attention from several experiments in many experimental physics application. SSCNs were demonstrated to perform better than a dense CNN for the task of semantic segmentation on an open data set sample~\cite{Domine:2019zhm}, as well as on background rejection using calibration data in a gas xenon TPC~\cite{Kekic:2020cne}. As the images produced by the MicroBooNE detector are very sparse macroscopically (after applying low frequency noise-filtering and signal-processing), but hold rich dense data in the vicinity of an interaction, the use of the SSCN adaptation, is appropriate. Moreover, as explained in Sec.~\ref{sec:net}, an SSCN's resource consumption scales linearly with the amount of data and thus is of much interest in future LArTPC detectors with a much larger data volume such as ICARUS~\cite{Amerio:2004ze} and DUNE~\cite{Acciarri:2016ooe} (for more examples of SSCN use in LArTPCs see~\cite{Domine:2020tlx,Koh:2020snv,Adams:2019uqx})

In this paper, we describe \sssnet\ (Sparse Semantic Segmentation Network), a deep-learning-based algorithm designed to distinguish showers from tracks in MicroBooNE. After applying a set of initial data selection criteria for reducing low energy backgrounds and tracks originating from cosmic muons, \sssnet\ is applied, replacing the previously used CNN~\cite{Adams:2018bvi}. The network is used for the task of semantic segmentation at the pixel level~\cite{long2014fully}, before any physical entities are identified (e.g., interaction vertex, grouping energy depositions originating from the same particle, etc.). It is worth mentioning that although \sssnet is currently used only within the DL-LEE analysis, It is not tuned to that specific analysis and can be used in the search of any type of search in a LArTPC. This study was done in a Singularity~\cite{Singularity} software container\footnote{https://singularity-hub.org/containers/6555}, and the implementation of \sssnet\ is available on GitHub\footnote{https://github.com/ranitay/SparseSSNet}.

\section{The Network}
\label{sec:net}

\sssnet treats each wire-plane separately, therefore three networks are trained one for each wire-plane. The architecture of the networks is similar however, the masking and the derived set of network weights are unique to each network. No data are shared between the different planes.

The main idea behind \sssnet\ is that, prior to training, masking is performed on the image to distinguish between important and non-important pixels, the masking is done on the feature vector (the pixel intensity this work). The input data therefore are reduced from $\mathrm{N} \times \vec{f}$ to $\mathrm{N}_{th} \times{d}\times\vec{f}$. N is the total number of pixels in the image and scales exponentially with the image dimension (N=$3456 \times 1008$ for images produced by MicroBooNE). $\vec{f}$ is the feature vector (the intensity of the pixel for the input layer in this analysis), $d$ is the dimension of the image ($d=2$ in this analysis), and $\mathrm{N}_{th}$ is the number of pixels passing selection criteria (masked). Using the sparse representation, the data stored, as well as the number of computations done for a convolutional layer, scales linearly with the number of pixels passing selection criteria instead of as a function of the total number of pixels.

\sssnet\ processes images using a sparse algebra as opposed to a dense one. Using \sssnet\ provides several benefits over using the dense CNN. The key benefit is the reduction in the time and memory consumption. The nature of the sparse representation drastically improves the time for processing a $512 \times 512$ image (this is the size of an image from the training sample see Sec.~\ref{sec:data}) from $\approx$\,5\,s to $\approx$\,0.5\,s (The performance tests were done on an Intel core i7-8750H CPU 2.2\,GHz). This improvement in the computation time reduces it below the time required of I/O which becomes the current bottleneck. In addition, as pixels of no interest (e.g., 0 intensity) are not saved, the memory consumption for inference is reduced from $\approx6$\,GB to $\approx1$\,GB. As a consequence, a full image can be inferred in a timely manner on a single CPU, unlike in the dense case where the images were cropped into $\approx64$ smaller images. Looking at the total CPU wall-time needed to analyze the current MicroBooNE open data set of $\approx 5 \times 10^19$ protons on target, a total of 195875 events (three planes each) requires $\approx2267$\,days for the dense case and only $\approx1.13$\,days in the sparse case (assuming a batch size of 1 as required by the dense case, due to CPU memory limitation). Due to these advantages, the image inference can be performed utilizing commonly available computing resources provided by numerous high throughput computing facilities.

Given the design of the MicroBooNE detector electronics, the pixel value (integrated counts over six TPC time ticks) distribution of minimum ionizing muons peaks at $\approx40$\,PIU and the maximal pixel value of a muon would produce a pixel value of $\approx220$\,PIU. In this analysis, all pixels with intensity smaller than $10$\,PIU or larger than $300$\,PIU are not included in the sparse representation. The lower value is set by the ability to distinguish signal from noise, while the higher removes un-physical noise originating in the data-processing and is set well above any expected value. This threshold reduces the number of relevant pixels to $\approx0.5\%$ of the total pixels in an image. 

\begin{figure}[ht]
\centerline{\includegraphics[width=1\linewidth]{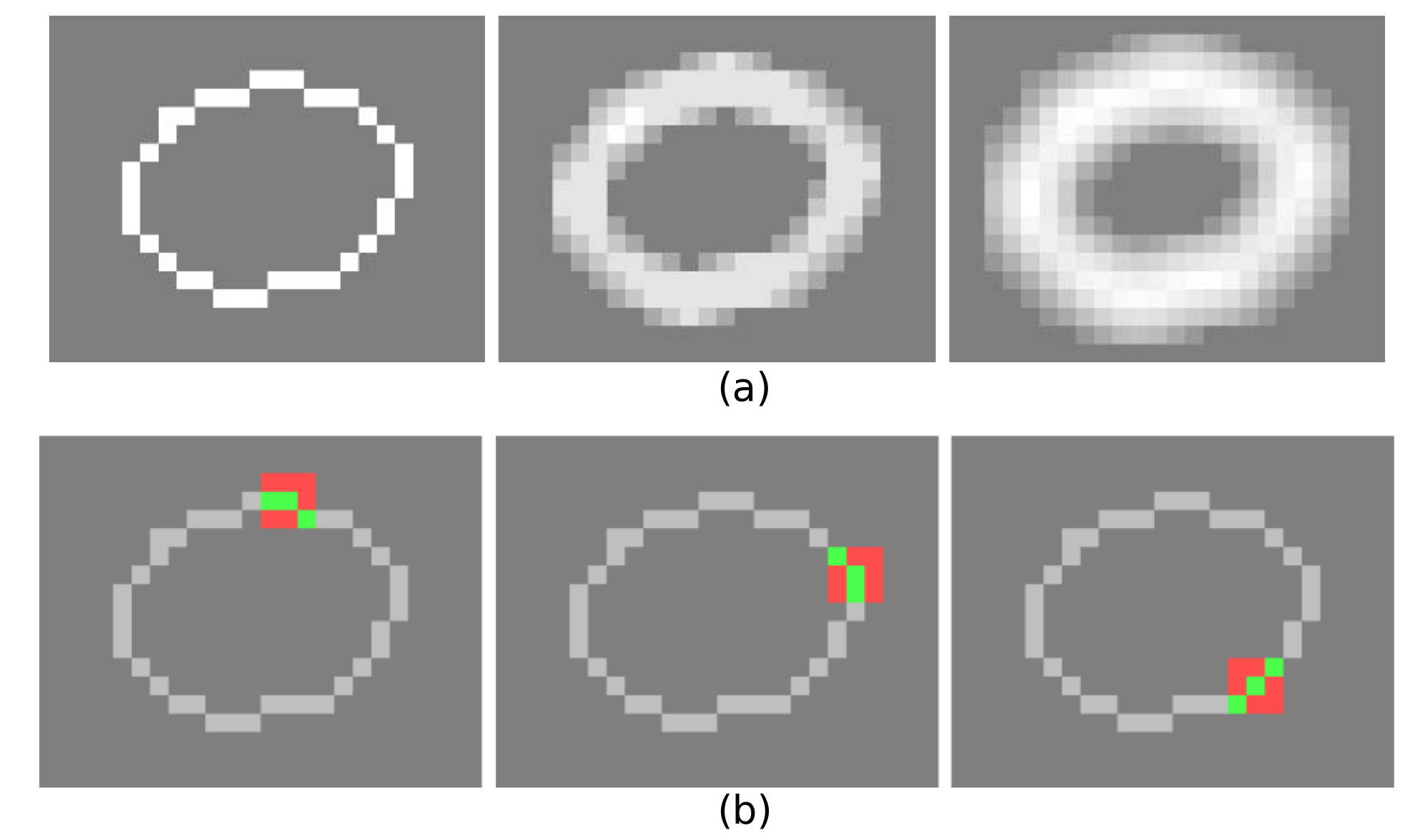}}
\caption{(a) an example of an image being dilated after two dense convolution operations using a filter with size 3\,$\times$\,3, weights 1\,/\,9, and stride 1. (b) a non-dilated image using sparse convolutional layers, the green label represents pixels that are kept for consecutive layers and the red label represents pixels that would have acquired values in dense CNNs, but do not in \sssnet\ (image taken from~\cite{Graham_2018_CVPR}).}
\label{fig:blur}
\end{figure}

Once a pixel is not included in the sparse representation, it is disregarded and not stored in the input data. In any convolution operation, an output pixel will be nonzero if and only if the central pixel of the receptive field is nonzero in the input feature map. Thus, the sparseness of the data is retained through a convolution, constraining only pixels which were non-zero in the input layer to be activated in hidden layers. While a dense convolution will spread information to pixels which originally contained no information (this process is referred to as ``image dilation'') the sparse convolution restricts changes to only those pixels which satisfy the selection criteria. This prevents image dilation (see Fig.~\ref{fig:blur}) and improves the accuracy of the network.

\begin{figure*}[!ht]
\centerline{\includegraphics[width=0.97\linewidth]{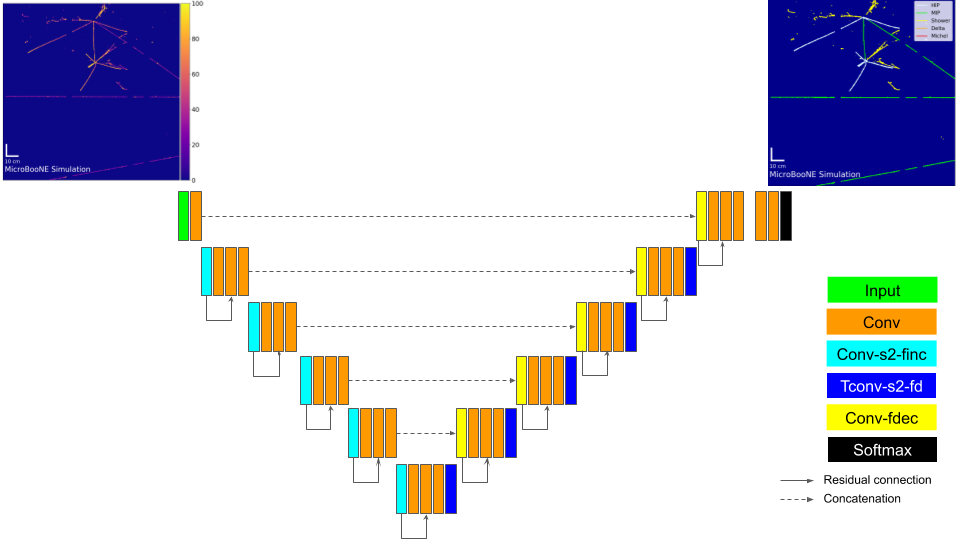}}
\caption{\sssnet's architecture. Light blue boxes represent convolutional layers with stride two (decreasing the spatial size) and an increased number of filters. Orange boxes represent convolutional operations. Dark blue boxes represent transpose convolution with stride two (increasing the spatial size) and a decreased number of filters. Yellow boxes are convolutional layers with stride one and decreased number of filters. The depth of the network is defined as five since there are five down-sampling operations. The spatial size is constant along the horizontal axis.} 
\label{fig:arch}
\end{figure*}

The architecture of \sssnet\ is U-Res-Net (see Fig.~\ref{fig:arch}) which is a hybrid of U-Net~\cite{10.1007/978-3-319-24574-4_28} and Res-Net~\cite{7780459}.
This network consists of two parts: an encoder and a decoder. At the encoder, the image is downsampled using a stride=2, and the network extracts features at various scales and hierarchical correlations. The number of downsampling steps is referred to as the depth of the network. The decoder upsamples the output of the encoder using transpose convolutions, and learns how to interpolate back to higher spatial resolution images, until reaching the original size. The feature map from the encoder is concatenated to the decoder feature map of the same size to help the decoder to restore the original image. Each block of convolutions (up or down sampling) is made of two convolutional layers followed by a batch normalization operation and a rectified linear unit (ReLU) function. Additional skip connections are added according to the Res-Net architecture.

\sssnet\ is constructed with 32 filters in the initial block and has a depth of five. The number of filters for a specific block is $i\times32$, where $i$ is the block's depth (the input layer has a depth of $1$ ). The filter size is $3\times3$ and a stride of two is used at each down-sample process, decreasing the image size by a factor of two. A softmax\footnote{Softmax is a mathematical function that takes as input a vector of real numbers, and maps it into a probabilities summed to one, with larger input values corresponding to higher probabilities.} classifier, an ADAM optimizer~\cite{Kingma2015AdamAM} and a cross-entropy loss function summed over all non-zero pixels are used. In addition, a pixel-weighting scheme is applied for preventing class imbalance (see Sec.~\ref{sec:weighting}).

The output of the network is a pixel-wise normalized five-dimensional probability vector $\vec{p}$ (also referred to as scores); the predicted pixel label is then defined to be the class with the highest probability.

\section{Simulated Data samples}
\label{sec:data}

\begin{figure}[p]
\centerline{\includegraphics[width=0.9\linewidth]{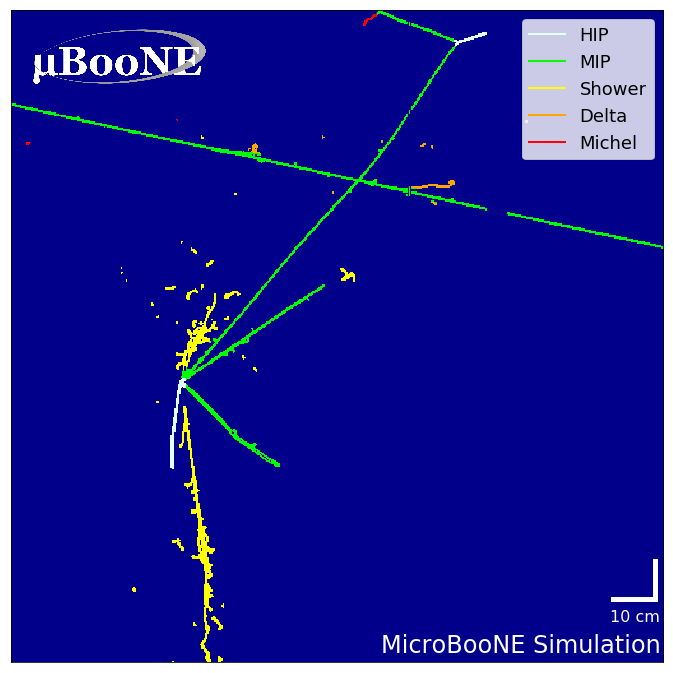}}
\caption{An example of a simulated event projected on the collection plane, taken from the training sample. Multiple particles are generated at a specific location and propagated throughout the detector to mimic a neutrino interaction. In addition a higher energy muon is simulated to mimic cosmic muons.}
\label{fig:simEx}
\end{figure}

The data samples consist of images that contain neutrino interactions, as well as many particles from cosmic rays, since the MicroBooNE detector runs on the surface and uses a long integration time to collect ionization over the long drift length. The neutrino interactions consist of electrons ($e$), photons ($\gamma$), muons ($\mu$), charged pions ($\pi^\pm$) and protons ($p$). Most of the cosmic ray particles are muons. These muons are higher in energy than the muons that are produced in neutrino interactions. Training is performed on simulated images consisting of the above mentioned particles in addition to higher energy muons~\ref{tab:samp}. An example of a simulated image with cosmic muons is shown in Fig.~\ref{fig:simEx}.

The training and test data samples consist of $\approx$\,120,000 and $\approx$\,23,000 simulated images respectively. For each image, pixel intensity and class labels (see~\ref{sec:lable}) are produced on which supervised learning is performed. For validation tests, several samples of $\approx$\,15,000 neutrino interactions assuming different variations on the detector response model are used.

Pixels are associated with the simulated particles that contributed to their intensity. All pixels associated with a given particle are grouped and considered a cluster for that particle (for a shower all pixels are considered part of the original particle (electron or gamma) initiating the shower). For each event, particle propagation as well as detector effects are applied to derive the final input image. The data samples are produced using the LArSoft~\cite{Snider:2017wjd} v08\textunderscore05\textunderscore00\textunderscore06 and UBOONECODE~\cite{uboone} v08\textunderscore00\textunderscore00\textunderscore13a packages. The simulation of the wire response is performed by the Wire-Cell simulation code, common to several current LArTPC experiments~\cite{Adams:2018gbi,Adams:2018dra}.

\subsection{Particle sample}
\label{sec:particle}

\sssnet is trained to be sensitive only to outgoing charged particles traveling in the LAr and producing ionization electrons (translated into PIUs); thus, the simulation sample contains only randomly produced charged particles and gammas with no assumption on the interaction that produced them. For each image in the training and test samples, a random location in the detector is drawn from a uniform distribution. A random number $N(=\Sigma n_i)$ of particles are generated at this location where $N$ is the total number of particles and $n_i$ is the number of particles from type $i$. $N$ is drawn from a uniform distribution in the range of 1--6, whereas $n_i$ is drawn from a uniform distribution in the ranges specified in Table~\ref{tab:samp}. The momentum vector direction of each particle is chosen from an isotropic distribution. Approximately $85\%$ of the sample contains particles with kinetic energies ($E_k$) consistent with neutrino interactions within MicroBooNE. The energy range (E) for this sample for each particle is shown in Table~\ref{tab:samp}. A smaller sample ($\approx$\,15\%) is generated with a different configuration targeted at low energy interactions where particle identification becomes more difficult. The momentum (P) range for each particle from the low energy (low E) sample is shown in Table~\ref{tab:samp}. The number of particles generated ($N$) and their multiplicity ($n_i$) remains the same. Finally, a random number of muons in the multiplicity range of 5--10 and kinetic energy range of (5,000--20,000)\,MeV are generated in both samples to mimic cosmic rays.

\begin{table*}[]
\centering
\caption{The data sample particle content. For each particle type the multiplicity per event, the kinetic energy range for the full sample, and the momentum for the low energy (E) sample are given. Notice that unlike the particles originating at the simulated ``interaction point'' the cosmic muons for the low E sample are still defined by their kinetic energy as they are the same for both samples.}
\label{tab:samp}
\resizebox{\textwidth}{!}{

\begin{tabular}{l||c c c c c c}
    \hline \hline
         Particle       & $e$    & $\gamma$ & $\mu$ & $\pi^\pm$  & p    & Cosmic $\mu$ \\
         \hline
         Multiplicity   & 0--2 & 0--2     & 0--2  & 0--2     & 0--3 & 5--10 \\
        $E_k$ [MeV]    & 50 -1,000 & 50--1,000 & 50--3,000 & 50-2,000 & 50--4,000 & 5,000--20,000 \\
         P (low E) [MeV/c]  & 30--100 & 30--100 & 85--175 & 95--195 & 300--450 & 5,000--20,000 ($E_k$) \\
    \hline \hline

\end{tabular}
}

\end{table*}

\subsection{Training labels}
\label{sec:lable}

In the preparation of the samples, training labels are assigned to each pixel according to the particle contributing to their intensity. A total of five different labels are used. For the training and test sets, a pixel can be a sum of ionization electrons produced by two different charged particles reaching the same wire at the same time. Therefore, we follow a one-hot label scheme, i.e., a pixel can have only one label. In the case a pixel can be assigned to more than one label, the label assigned to it will be the highest according to the following order. This improves the performance of downstream analysis tasks such as identifying vertices~\cite{Abratenko:2020wum}. 

\begin{enumerate}
    \item \textbf{heavily ionizing particles (HIP)}, produced by protons, typically manifest in a short, highly ionized track.
    \item \textbf{minimum ionizing particles (MIP)}, produced by muons and charged pions, typically manifest in a longer, fainter (lower dE/dx) track. 
    \item \textbf{Showers}, produced by electrons, positrons, and photons above a minimal energy, $\approx$\,39 MeV in liquid argon~\cite{Adams:2019law}.
    \item \textbf{Delta rays}, produced from ejected atomic electrons from a hard scattering of other charged particles, mainly muons.
    \item \textbf{Michel electrons}, produced from a decay at rest of muons.
\end{enumerate}

Within the current DL-LEE analysis, only two classes are used for particle ID, track and shower; thus, the previously mentioned five-classes are mapped into two new classes. 

\begin{enumerate}
    \item \textbf{Track}, either HIP or MIP labels.
    \item \textbf{Shower}, either shower or delta ray or Michel electron labels.
\end{enumerate}

Notice that the re-classification does not require new inferring; rather it is just a mapping of these original five labels to the newly defined two labels (i.e., if a pixel is classified by the network as a MIP in the DL-LEE analysis it will be considered as a track). We intend to exploit the full class feature set in future analyses.

\subsection{Full neutrino interaction sample}
\label{sec:nusimsample}

The neutrino simulation is performed in two stages in order to model the cosmic background in a more realistic manner. The first stage is simulating the neutrino interaction itself using the GENIE ~\cite{Andreopoulos:2015wxa} v3\textunderscore00\textunderscore04 and Geant4~\cite{Agostinelli:2002hh} v4\textunderscore10\textunderscore3\textunderscore03c software packages in addition to the LArSoft  and UBOONECODE packages. The second is using a sample of beam-off data, taken from beam-off periods triggered on cosmic rays, and overlaying it on the neutrino interaction. As particles from the beam-off data sample are not from simulation, no labels can be assigned to them. 

There are two Monte Carlo (MC) simulation samples used for the study. The full-BNB sample includes all types of neutrino interactions that are expected to occur in MicroBooNE. The intrinsic $\nu_e$ sample comprises events due to the electron-flavor neutrinos predicted to be in the flux, with all neutrino interaction types included. Both samples are overlayed with beam-off data.

\textsc{SparseSSnet}'s predictions may vary depending on several factors, but the primary variation is from modeling different aspects of the detector response such as electric field, space-charge effects~\cite{Abratenko:2020bbx}, wire response, etc. Therefore to assess systematic uncertainties (see Sec.~\ref{sec:nusim}) samples with variation in the detector response model are generated with O(15,000) events per each of the eight variations explored. The sample with the nominal detector response is referred to as the central value simulation sample. These samples are used to verify the performance of \sssnet\ within the context of the DL-LEE analysis; therefore, only the two-class semantic segmentation is studied.

\section{Pixel Weighting}
\label{sec:weighting}

To prevent class imbalance~\cite{876973720021001}, a case where one class dominates the loss function and the penalty for incorrect prediction of other classes is negligible, we apply a pixel weighting scheme, defining the loss function

\begin{equation}
    \textrm{Loss} = \Sigma_i w_i \cdot (\vec{l}_i \cdot \textrm{log}(\vec{P}_i))
\end{equation}
where $w_i$ is the weight defined for each pixel, $\vec{l}_i$ is the label vector of pixel $i$ (e.g., (1,0,0,0,0) for a HIP) and $\vec{P}_i$ is the scores vector for pixel $i$ (e.g., (0.8,0.2,0,0,0) for 80\% HIP and 20\% MIP).

\begin{figure}[!htb]
\centerline{\includegraphics[width=0.8\linewidth]{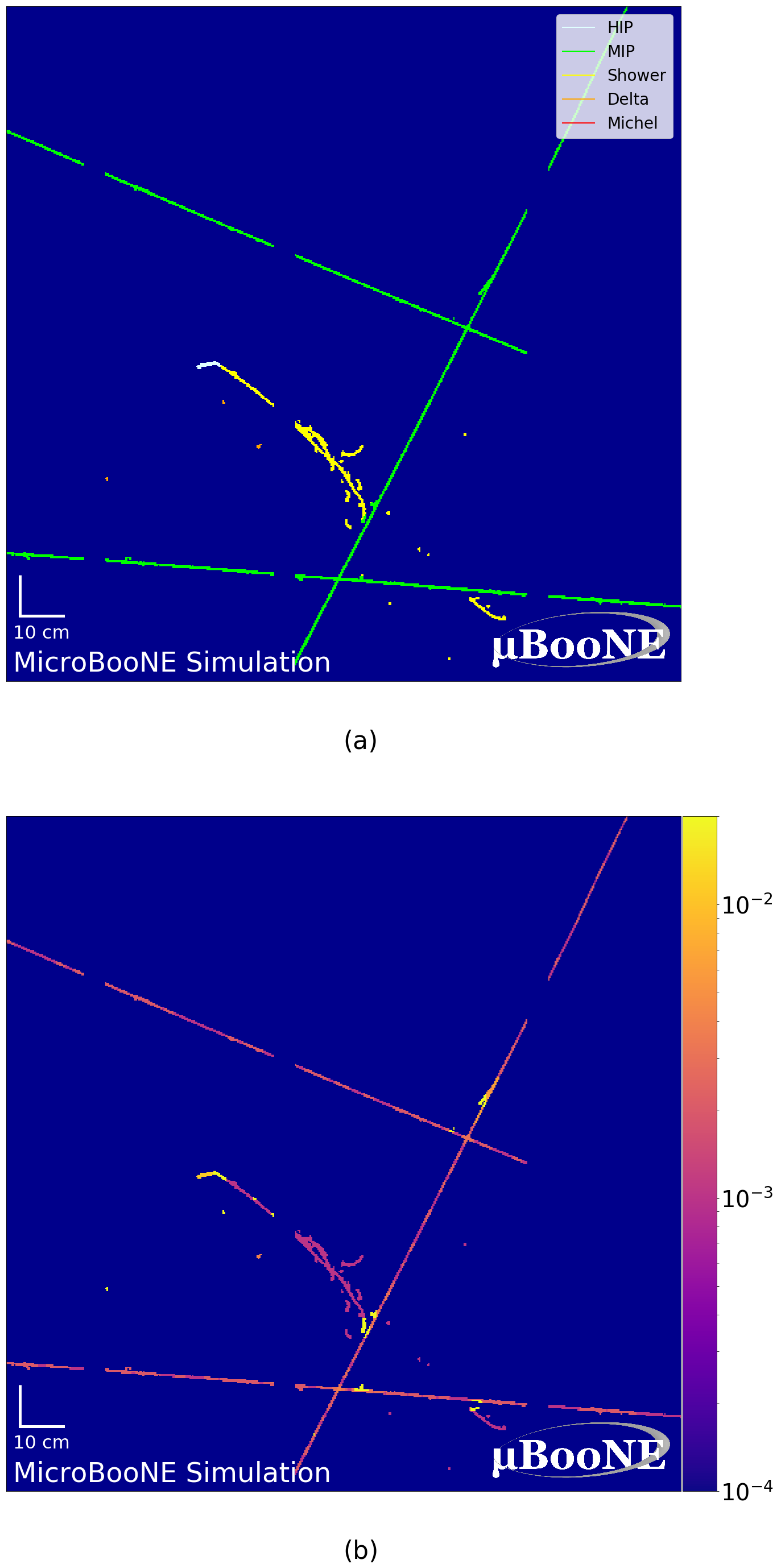}}
\caption{An example of a simulated event projected on the collection plane, taken from the training sample. (a) The labels assigned for each pixel according to generated particles. (b) Pixel weighting: for each cluster a weight proportional to the inverse of its size is assigned in the range of $(0.02$ -- $2)\times 10^{-2}$. Notice that even though the shower has gaps in it (due to gammas) all pixels of the shower are associated with the original electron cluster for the purpose of cluster weighting. For crossing type pixels a weight of $2 \times 10^{-2}$ is assigned.}
\label{fig:weight}
\end{figure}

The sum of two types of weighting is assigned to each pixel: \textbf{cluster weighting} and \textbf{vertex weighting} (see Fig.~\ref{fig:weight}).
\begin{itemize}
    \item \textbf{Cluster weighting:} large clusters contain many pixels. This makes it easier to correctly label them, as they contain more information. Moreover, labeling a large cluster correctly reduces the loss function by a significant amount (proportional to the number of pixels in the cluster). Therefore, small clusters should be treated with more care to prevent the loss function from being governed by one correctly labeled big cluster. We apply a cluster weight that is inversely proportional to the size of the cluster in the range of $(0.02$ -- $2)\times 10^{-2}$.
    \item \textbf{Vertex Weighting:} pixels at the center of a cluster are easier to identify, as they cannot be confused by other pixels in their close vicinity. On the other hand, pixels near a cluster labeled as a ``different class'' are the most difficult to recognize and impact the vertex reconstruction algorithm~\cite{Abratenko:2020wum} (the next stage of the DL-LEE analysis) dramatically (e.g., near a vertex of a proton and a muon it is harder to distinguish exactly which pixel is associated with the proton and which with the muon). We apply a vertex weight of 0.02 to pixels within a three-pixel distance from a pixel from a different class, and vertex weight of zero to all other pixels.
\end{itemize}

\section{Results}
\label{sec:results}

\subsection{Test sample}

For each plane, \sssnet\ is trained for $\approx$\,15 epochs\footnote{An epoch refers to one cycle through the entire data set. In this analysis, as images are selected randomly for each batch, an epoch refers to a cycle through a number of images equal to the sample size ($\approx$\,120,000 images).}, the chosen weights are obtained from $\approx$\,8.5 epochs of training to achieve the best performance without over fitting the network. Although the performance of the signal processing is plane dependent~\cite{Adams:2018dra}, the \sssnet\ results from all planes are similar (averaged over the entire phase-space mentioned in table~\ref{tab:samp}) and therefore we discuss only results obtained from the collection plane. We define the accuracy as the number of correctly classified pixels with respect to non-zero pixels only.

\begin{figure}[t]%
 \centering
 \centerline{\includegraphics[width=1\linewidth]{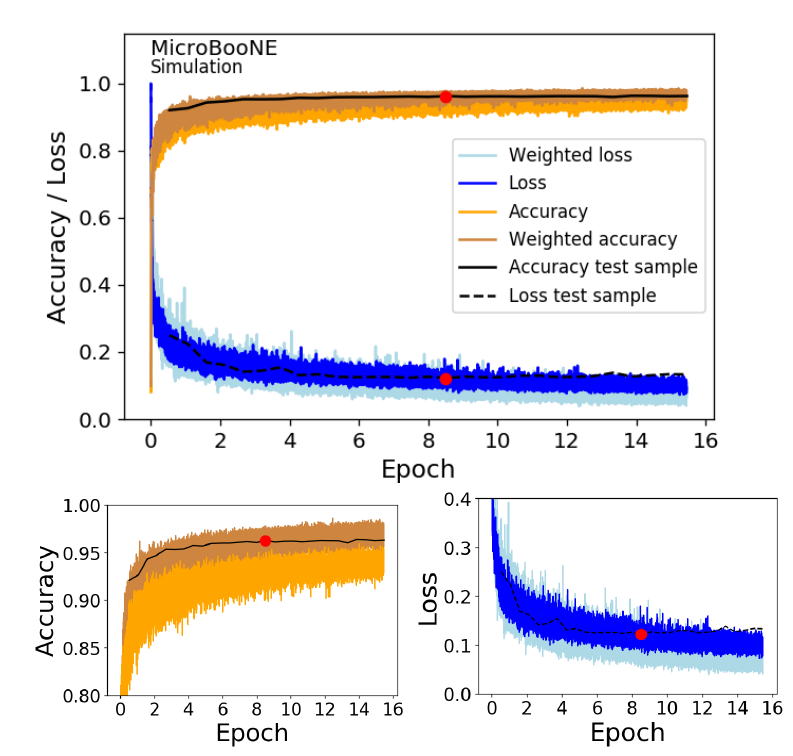}}

 \caption{ (Top) The accuracy and loss of the network on the training and inference data sets obtained from the collection plane. The accuracy before (orange) and after (light brown) applying weighting, the loss function normalized by the loss after the first iteration, before (blue) and after (light blue) applying weighting. The accuracy of the inference data sample is indicated in solid black and the loss on the test sample is indicated in dashed black. The selected network weights are indicated by the red dot (8.5 epochs). (Bottom) plots are zoomed in.}%
 \label{fig:AccLoss}%
\end{figure}

The total accuracy obtained from the collection plane test sample is 96\% for the final configuration of the network. The accuracy and the loss function obtained from the training sample both with and without weighting are shown in Fig.~\ref{fig:AccLoss}, along with the accuracy and loss obtained from the test sample. Applying the pixel weighting improves the total accuracy from $92\%$ to $96\%$. 

\begin{figure}[htb]
\centerline{\includegraphics[width=0.95\linewidth]{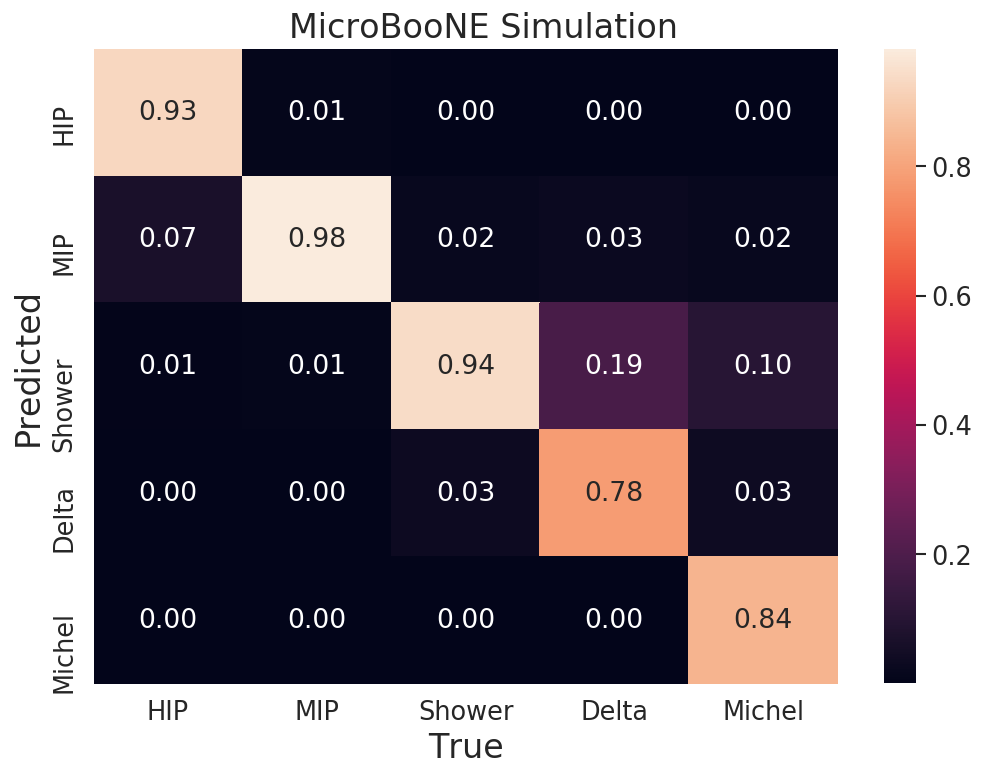}}
\caption{ Five-class confusion matrix obtained from the collection plane test sample. Each box represents the fraction of pixels that are from the class stated in the x-axis and predicted as class stated in y-axis from the test sample. The smallest number of pixels is O($10^5$) for Michel electrons. All other classes vary between $10^6-10^7$ pixels.}
\label{fig:Conf5Y}
\end{figure}

A better quantification of the performance with respect to each class is achieved by the confusion matrices. The five-class semantic segmentation confusion matrix obtained from the test sample for the collection plane is shown in Fig.~\ref{fig:Conf5Y}. 

The number of pixels obtained for each class is O($10^5$) for Michel electrons, and varies between  $10^{6}$ and $10^{7}$ for other classes; thus statistical uncertainties are small. The matrices for the induction planes are fairly similar and are presented in appendix~\ref{app:conf}.

In the DL-LEE analysis, only two classes are defined (see Sec.~\ref{sec:lable}). The accuracy for each class for the two-class semantic segmentation obtained for the collection plane (Y) is shown in Table~\ref{tab:mat2}. The results from the two induction planes are similar and are presented in appendix~\ref{app:conf}. The number of pixels obtained for each class is $>10^7$; as with the case of the five-class scheme, the statistical uncertainties are negligible. Comparing these results to the previous network used~\cite{Adams:2018bvi}, yields an improvement in the shower accuracy from 95.9\% to 99.6\% and in the track accuracy from 97.4\% to 99.2\%. The accuracy improvement is due to the no-image-dilation effect, the lack of background class, and to fewer boundaries in the image (all zero and non-zero pixels)

\begin{table}[]
\centering
\caption{\sssnet's track and shower accuracy, for the test sample and the neutrino interaction central value simulation sample (both full-BNB and intrinsic $\nu_e$). Results are obtained from the collection plane. The number of pixels associated with each class is O($10^7$) pixels except for the full-BNB shower which is O($10^5$). The drop in the shower accuracy for the neutrino interaction sample is explained in Sec.~\ref{sec:nusim}.}
\label{tab:mat2}
\resizebox{0.4\textwidth}{!}{

\begin{tabular}{l||ccc}
\hline \hline 

       & Test & Intrinsic $\nu_e$ & Full-BNB \\
       \hline
Track  & 0.992 & 0.992   & 0.998  \\
Shower & 0.996 & 0.859   & 0.823 \\ \hline \hline
\end{tabular}
}

\end{table}

\begin{figure*}[!htb]
\centerline{\includegraphics[width=\linewidth]{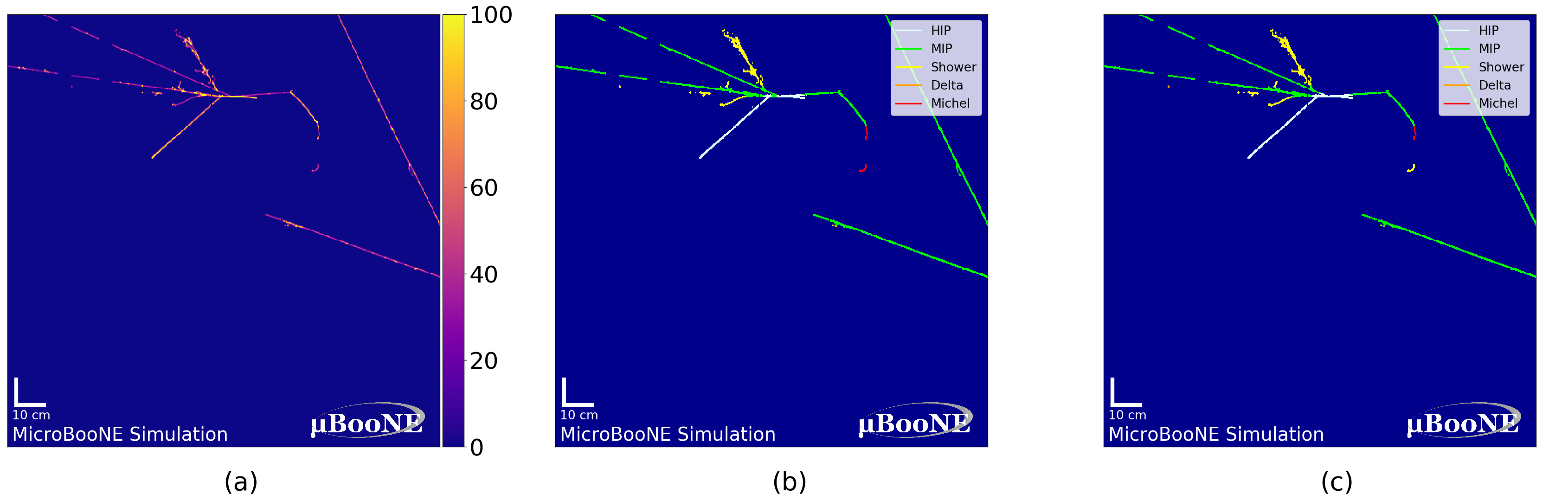}}
\caption{An example of a simulated event from the test sample projected on the collection plane. (a) pixel intensity. (b) Truth label. (c) \sssnet\ predictions. All images are 512~$\times$~512 pixels crops from a full detector simulation.} 
\label{fig:myMC19699}
\end{figure*}

An example of an event from the test sample is presented in Fig.~\ref{fig:myMC19699}. This display encapsulates the performance of the network and contains the pixel intensity, the truth level label, and the \sssnet's predictions.

\subsection{Neutrino interaction sample}
\label{sec:nusim}

The neutrino interaction sample contains beam-off data which are not assigned with labels (see Sec.~\ref{sec:nusimsample}). Due to labeling priorities (see Sec.~\ref{sec:lable}), when a cosmic muon crosses a shower, \sssnet\ is trained to predict the joint pixels as MIPs and not as showers. These pixels are considered as wrongly predicted which biases the network's shower accuracy. The accuracy calculated for the two-class semantic segmentation task for both the full-BNB and intrinsic $\nu_e$ samples are shown in table.~\ref{tab:mat2}. The central value simulation samples are used to calculate these results. 

\begin{figure*}[!hb]
\centerline{\includegraphics[width=\linewidth]{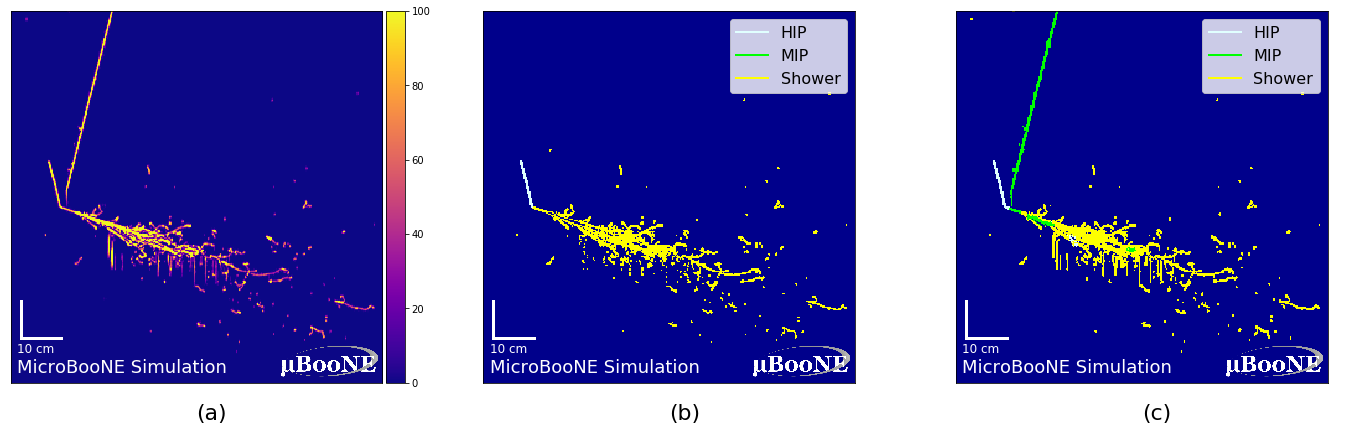}}
\caption{An example of a simulated $\nu_e$ interaction projected on the first induction plane. This is a (300~$\times$~300)\,pixels crop from the full detector image shown in Fig.~\ref{fig:nue1}. (a) Pixel intensity of interaction overlayed with cosmic rays. (b) The produced particles upon interaction, before overlaying cosmic rays. (c) \sssnet\ predictions. Notice that although some shower pixels are mis-classified, the fraction of shower-like pixels is larger than 0.2 and therefore this interaction will be correctly classified as $\nu_e$ interaction (see Sec.~\ref{sec:sys}).  }
\label{fig:nue1z}
\end{figure*}

\begin{figure*}[]
\centerline{\includegraphics[width=\linewidth]{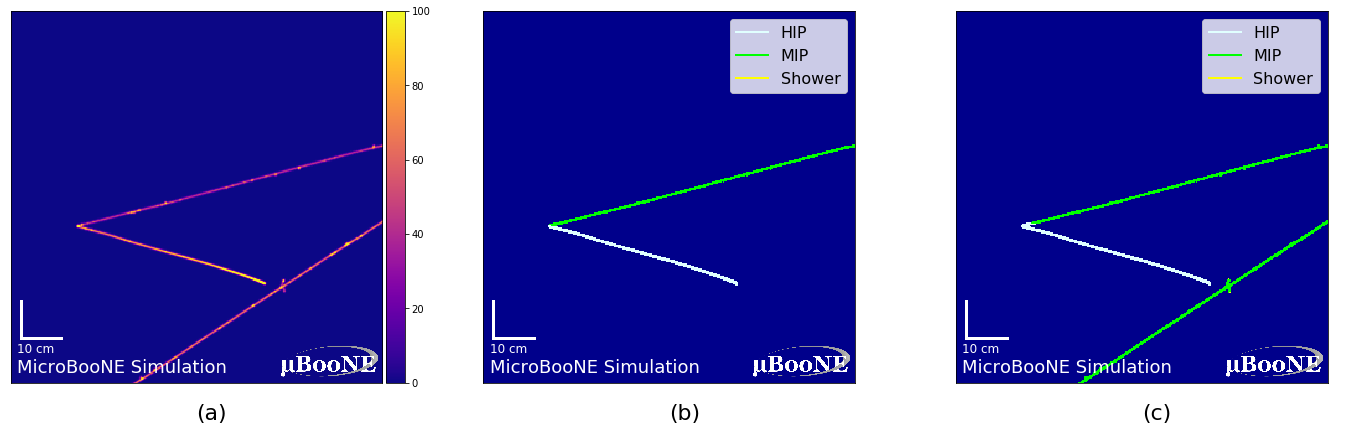}}
\caption{An example of a simulated $\nu_\mu$ interaction projected on the collection plane. This is a (300~$\times$~300)\,pixels crop from the full detector image shown in Fig.~\ref{fig:numu1}. (a) Pixel intensity of interaction overlayed with cosmic rays. (b) The produced particles upon interaction, before overlaying cosmic rays. (c) \sssnet\ predictions.}
\label{fig:numu1z}
\end{figure*}

The track accuracy is comparable to the one calculated from the test sample. The lower shower accuracy is attributed to the bias explained above. The track accuracy compared to previous analysis yields an improvement from 95.7\% to 99.8\% and 86.2\% to 99.2\% for the full-BNB (compared with the $\nu_\mu$ sample) and intrinsic $\nu_e$ (compared with the $\nu_e$ sample), respectively. Notice that the previous analysis did not use beam-off data and did not see the bias in the shower accuracy. Examples of simulated $\nu_e$ and $\nu_\mu$ from the neutrino interaction sample are shown in Fig.~\ref{fig:nue1z} and Fig.~\ref{fig:numu1z} respectively. These images are $300\times300$ pixels crops, roughly centered at the interaction point. For the full detector images see appendix~\ref{app:nuInter}. Each figure consists of the pixel intensity of the generated interaction with beam-off data, the generated interaction particles labels, and the \sssnet's predictions.

\subsection{Beam data}
\label{sec:real_data}
\begin{figure*}[!htb]
\centerline{\includegraphics[width=0.8\linewidth]{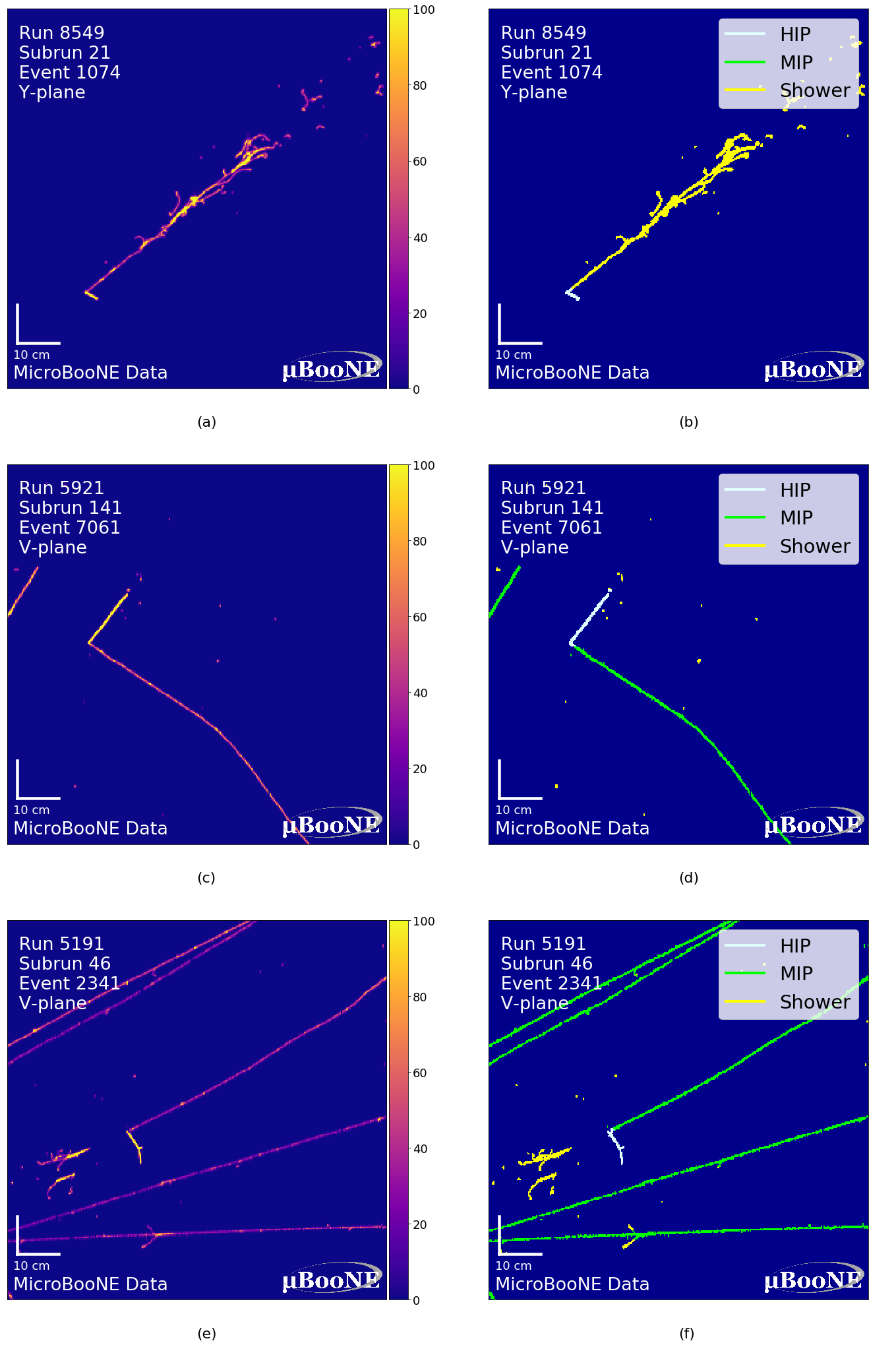}}
\caption{ Selected example from BNB data, image cropped to(300~$\times$~300)\,pixels roughly around the interaction point. The left column is the pixel intensity (a,c,e) and the right column is \sssnet\ predictions. The top (a,b) event is an example of a $1e1p$ final state topology, the middle (c,d) is an example of a $1\mu1p$ final state topology, and the bottom (e,f) is an example of a $1\mu1p1\pi^0$ final state topology. These events show good performance of \sssnet\ on data}
\label{fig:BNBdata}
\end{figure*}

As MicroBooNE has revealed part of the collected data, a final check of the performance of the network can be done on real BNB data. To quantitatively asses the performance of the network on BNB data, a more through analysis, including reconstruction and selection needs to be performed which is beyond the scope of this paper. However, we have hand scanned many events (>1000 selected 1l1p events) and saw no strange bias. We present example event-displays showing \sssnet\ predictions. The three selected events are presented in Fig.~\ref{fig:BNBdata}, the left column is the PIU which is the input to the network and right column is \sssnet\ prediction. The events themselves present a $1e1p$ (a and b), $1\mu1p$ (c and d), and $1\mu1p1\pi^0$ (e and f) final states.

\subsection{Robustness of results}
\label{sec:sys}

The output of \sssnet\ is used within the DL-LEE analysis in two tasks: pixel classification and cluster classification. The samples with variation in the detector model (see~\ref{sec:nusimsample}) are used to verify that any mismodeling of the detector does not strongly affect these classification tasks.

The pixel classification is used for vertex finding, shower reconstruction and energy estimation, cluster classification, and more (ongoing work). It is performed by setting a threshold on the shower score $p_{shower} \geq 0.5$. The amount of misclassified shower/track pixels have a $\approx$\,0.5\% variation between all different simulation models. 

The cluster classification is used to distinguish between a $\nu_e$ CCQE interaction ($1e1p$) and a $\nu_\mu$ CCQE interaction (1$\mu1p$) by applying a selection criterion of $f_s \geq 0.2$ for at least one cluster, where $f_s$ is the fraction of shower-like pixels in a cluster. The variations on shower clusters are $\approx$\,1\%. 

As discussed previously, the systematic errors are evaluated by running \sssnet\ on samples with detector variations. We find that, for the DL-LEE analysis, the uncertainty coming from \sssnet\ is negligible compared to uncertainties from the traditional algorithms used for track/shower reconstruction (ongoing work).

\section{Summary}
We have presented the performance of \sssnet\ in the task of semantic segmentation on simulated data from the MicroBooNE detector. The output of \sssnet\ plays an important role in many tasks such as vertex finding, shower reconstruction and energy reconstruction, and neutrino selection in the DL-LEE search in MicroBooNE, and is the first usage of SSCN in LArTPC data (beam-off) and realistic MC (neutrino interactions). The adaptation to sparse representation dramatically improves the inference time from $\approx$\,5\,s to $\approx$\,0.5\,s as well as the memory usage from $\approx$\,5\,GB to $\approx$\,1\,GB. In addition, there is an improvement in the accuracy of the test sample due to no-image-dilation preserving the sparsity and locality of the information. The current analysis uses only two classes (track and shower), however the network produces five-class segmentation which we plan to exploit in future analyses. 

\FloatBarrier

 The method that we have described here is transferable to other LArTPC detectors. This includes ICARUS, SBND, that are about to begin running on the same neutrino beamline as MicroBooNE, as well as for the DUNE experiment. 


\section{Acknowledgement}
This document was prepared by the MicroBooNE collaboration using the resources of the Fermi National Accelerator Laboratory (Fermilab), a U.S. Department of Energy, Office of Science, HEP User Facility.Fermilab is managed by Fermi Research Alliance, LLC (FRA), acting
under Contract No. DE-AC02-07CH11359. MicroBooNE is supported by the following: the U.S. Department of Energy, Office of Science, Offices of High Energy Physics and Nuclear Physics; the U.S. National Science 
Foundation; the Swiss National Science Foundation; the Science and Technology Facilities Council (STFC), part of the United Kingdom Research and Innovation; and The Royal Society (United Kingdom). Additional support for the laser calibration system and cosmic ray tagger was provided by the Albert Einstein Center for Fundamental Physics, Bern, Switzerland.

\appendix
\section{RESULTS FROM INDUCTION PLANES} 
\label{app:conf}
\begin{figure}[!htb]
\centerline{\includegraphics[width=0.93\linewidth]{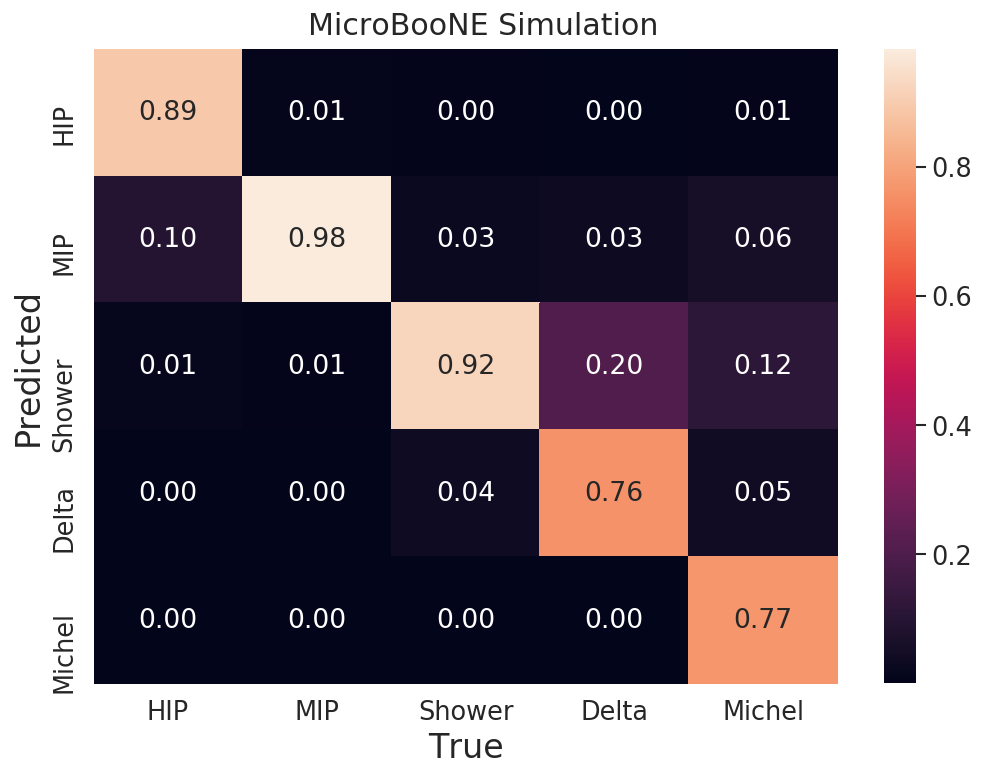}}
\caption{Five-class confusion matrices for the first induction  plane test sample. Each box represents the fraction of pixels which are from the class stated on x-axis and predicted as class stated in y-axis. The smallest number of pixels is O($10^5$) for Michel electrons. All other classes vary between $10^6-10^7$ pixels.}
\label{fig:Conf5U}
\end{figure}

\begin{figure}[!htb]
\centerline{\includegraphics[width=0.93\linewidth]{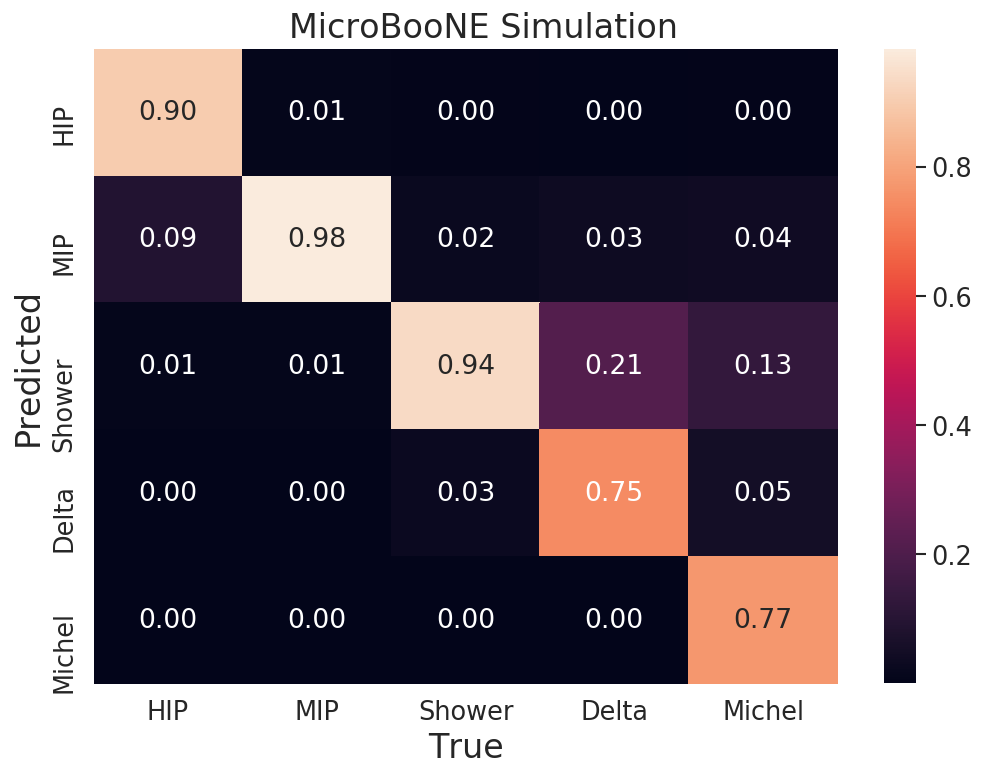}}
\caption{Five-class confusion matrices for the second induction plane test sample. Each box represents the fraction of pixels which are from the class stated on x-axis and predicted as class stated in y-axis. The smallest number of pixels is O($10^5$) for Michel electrons. All other classes vary between $10^6-10^7$ pixels.}
\label{fig:Conf5V}
\end{figure}

The five-class semantic segmentation confusion matrices produced from the data samples from the U plane (Fig.~\ref{fig:Conf5U}) and V plane (Fig.~\ref{fig:Conf5V}) are similar to the one presented in Fig.~\ref{fig:Conf5Y}, and are presented here for completeness. Notice that these matrices are obtained from different planes which yields different images (though coming from same interactions) and different networks; this can explain the small accuracy differences between the planes. The number of pixels obtained for each class is similar to the collection plane and is O($10^5$) for Michel electrons, and varies between  $10^{6}$ and $10^{7}$ for other classes; thus statistical uncertainties are small.

The accuracy for each class for the two-class semantic segmentation task obtained from the two induction planes on all simulation samples are similar to the those presented in Sec.~\ref{tab:mat2} and presented in table~\ref{tab:mat2_app} for completeness. The drop in the shower accuracy for the full-BNB and intrinsic $\nu_e$ samples is explained in Sec.~\ref{sec:nusim}. 
\\
\begin{table}[!ht]
\centering
\caption{\sssnet\ track and shower accuracy for the test sample and the neutrino interaction central value simulation samples (both full-BNB and intrinsic $\nu_e$). The results are obtained from the two induction planes. The number of pixels associated with each class is O($10^7$) pixels except for the full-BNB shower which is O($10^5$). The drop in the shower accuracy for the neutrino interaction sample is explained in ~\ref{sec:nusim}.}
\label{tab:mat2_app}
\resizebox{0.45\textwidth}{!}{
\begin{tabular}{l||ccccccc}
\hline \hline 

       & \multicolumn{2}{c}{Test} & \multicolumn{2}{c}{Intrinsic $\nu_e$} & \multicolumn{2}{c}{Full-BNB} \\
       & U       & V      & U     &V        &U      & V      \\ \hline
Track  & 0.988   & 0.990   & 0.990 & 0.989   & 0.997  & 0.998  \\
Shower & 0.996   & 0.994  & 0.823 & 0.858   & 0.809 & 0.821  \\ \hline \hline

\end{tabular}
}

\end{table}

\section{NEUTRINO INTERACTION EVENT DISPLAYS}
\label{app:nuInter}
\begin{figure*}[!b]
\centerline{\includegraphics[width=1\linewidth]{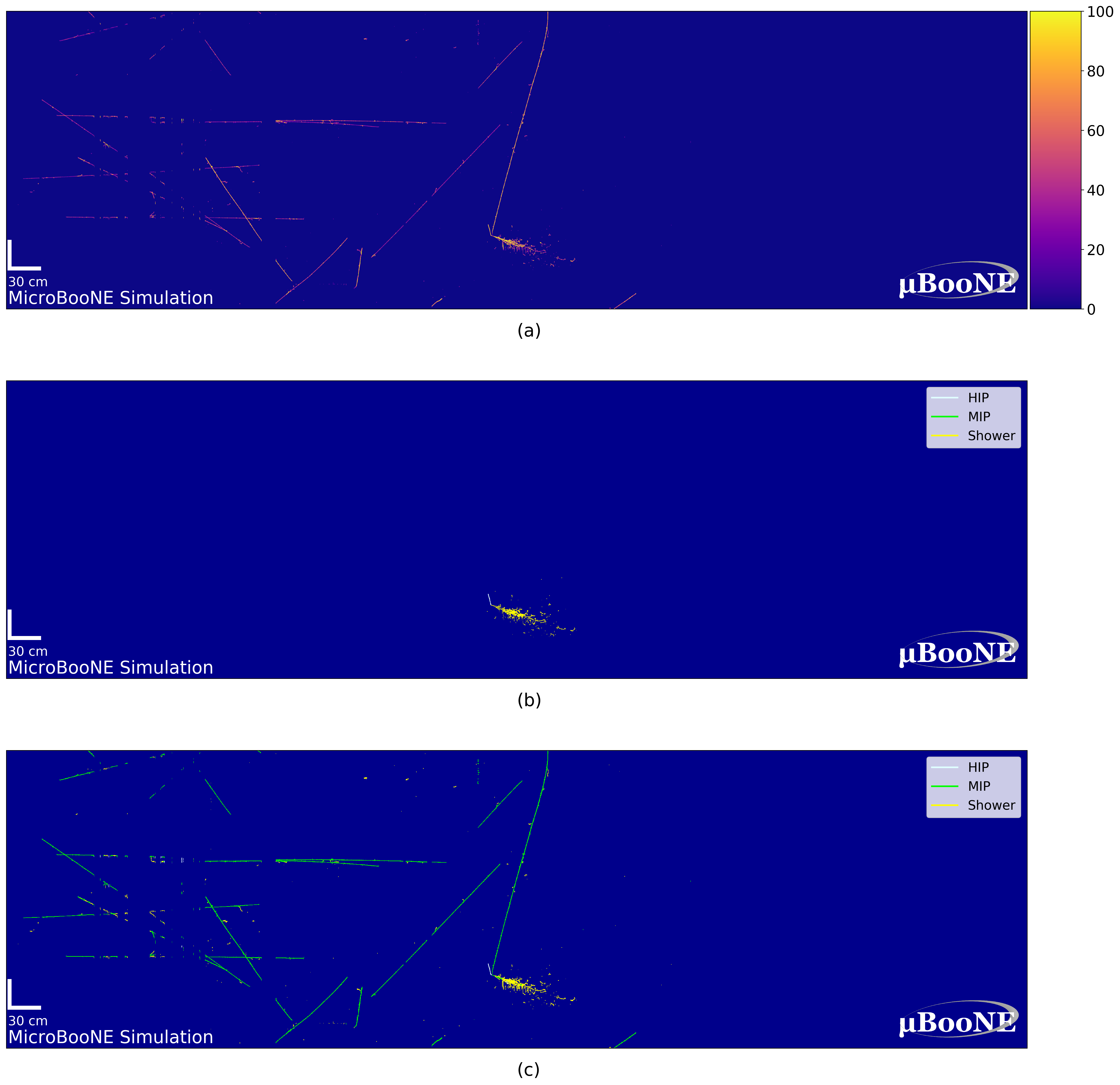}}
\caption{An example of a simulated $\nu_e$ interaction projected on the first induction plane. This is the full detector image of the $\nu_e$ interaction shown in Fig.~\ref{fig:nue1z}. (a) Pixel intensity of interaction overlayed with cosmic rays. (b) The label assigned to the simulated neutrino interaction. As cosmic rays are taken from beam-off data they are not assigned with labels. (c) \sssnet's predictions. }
\label{fig:nue1}
\end{figure*}

\begin{figure*}[!ht]
\centerline{\includegraphics[width=1\linewidth]{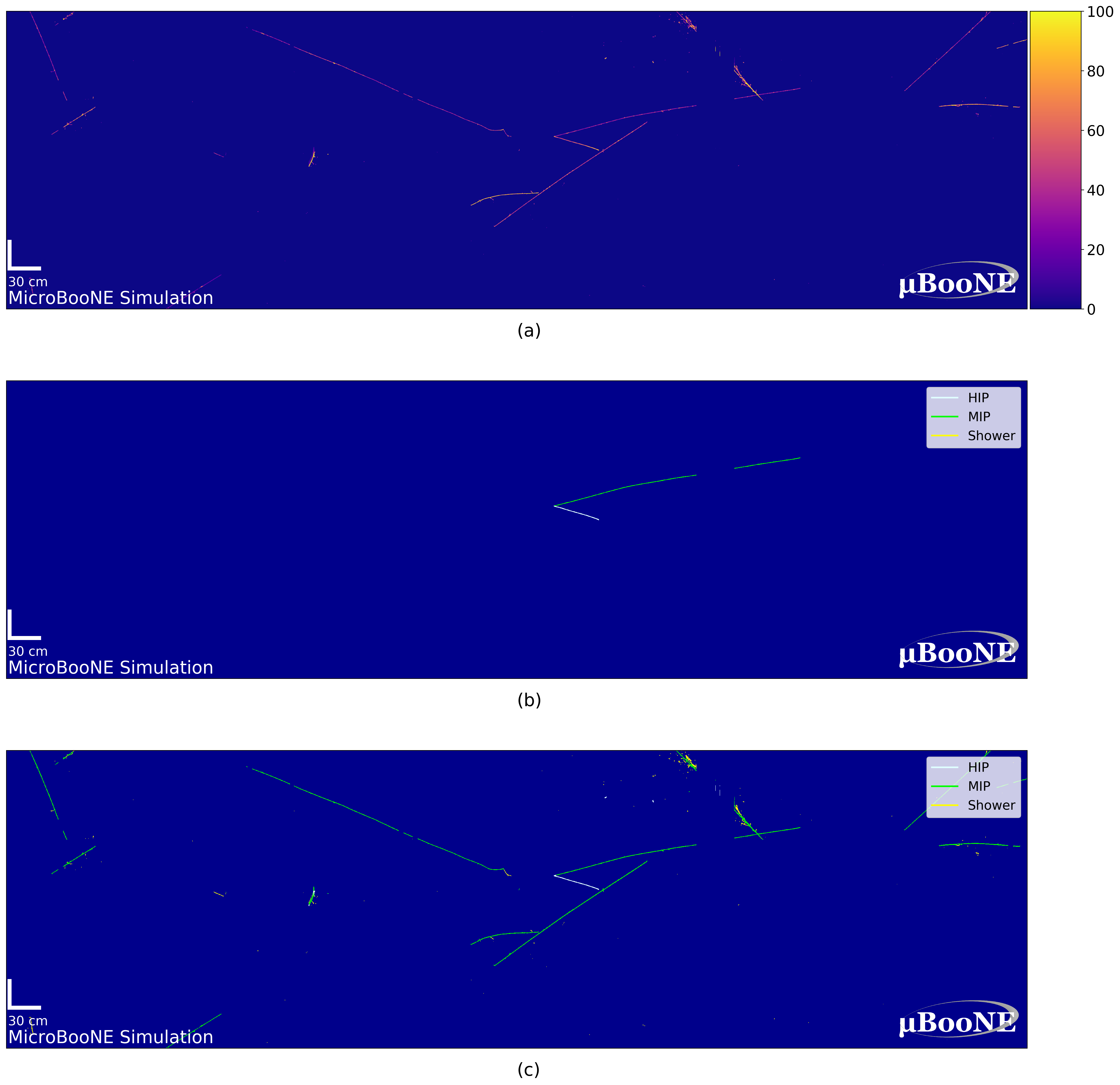}}
\caption{An example of a simulated $\nu_\mu$ interaction projected on the first induction plane. This is the full detector image of the $\nu_e$ interaction shown in Fig.~\ref{fig:numu1z}. (a) Pixel intensity of interaction overlayed with cosmic rays. (b) The label assigned to the simulated neutrino interaction. As cosmic rays are taken from beam-off data they are not assigned with labels. (c) \sssnet's predictions.}
\label{fig:numu1}
\end{figure*}

Event displays of $\nu_e$ (Fig.~\ref{fig:nue1}) and $\nu_\mu$ (Fig.~\ref{fig:numu1}) interactions from the full detector (3456~$\times$~1008 pixels) of the events shown in Sec.~\ref{sec:nusim}. Notice that induction planes (U \& V) contains only 2400 wires; hence, to keep the size of the images identical, an additional 1056 (2400 - 3456) are added but are empty. One can also see the sparsity of events from these figures.

%

\end{document}